\documentclass{PoS}
\usepackage{enumerate}
\usepackage{amssymb}
\newcommand{\Bs}{\mathrm{B}_\mathrm{s}}
\newcommand{\B}{\mathrm{B}}
\newcommand{\K}{\mathrm{K}}
\newcommand{\uq}{\mathrm{u}}
\newcommand{\bq}{\mathrm{b}}
\newcommand{\hq}{\mathrm{h}}
\newcommand{\Vo}{\mathrm{V}_\mathrm{0}} 
\newcommand{\Ao}{\mathrm{A}_\mathrm{0}} 
\newcommand{\Vk}{\mathrm{V}_\mathrm{k}} 
\newcommand{\beq}{\begin{equation}}
\newcommand{\eeq}{\end{equation}}
\newcommand{\beqa}{\begin{eqnarray}}
\newcommand{\eeqa}{\end{eqnarray}}
\newcommand{\Vub}{\left| V_\mathrm{ub} \right|}
\newcommand{\stat}{\mathrm{stat}}
\newcommand{\RGI}{\mathrm{RGI}}
\newcommand{\sRGI}{\mathrm{stat,RGI}}
\newcommand{\MSb}{\overline{\mathrm{MS}}}

\title{Form factors in the $\Bs \rightarrow \K \ell \nu$ decays using HQET and the lattice}

\ShortTitle{Form factors in the $\Bs \rightarrow \K \ell \nu$ decays using HQET and the lattice}

\author{\speaker{Debasish Banerjee}\\
       John von Neumann Institute for Computing (NIC), DESY, Platanenallee~6, 15738 Zeuthen, Germany.\\
       for the ALPHA collaboration. \\
       E-mail: \email{debasish.banerjee@desy.de}
 }

\abstract{We report on a recent computation of the 
  form factors in semi-leptonic decays of the $\Bs$ 
  using Heavy Quark Effective Theory (HQET) formalism 
  applied on the lattice. The connection of the form factors 
  with the 2-point and 3-point correlators on the lattice is 
  explained, and the subsequent non-perturbative renormalization 
  of HQET and it's matching to $N_f=2$ QCD is outlined. The 
  results of the (static) leading-order calculation in the 
  continuum limit is presented.}

\FullConference{34th annual International Symposium on Lattice Field Theory\\
         24-30 July 2016\\
         University of Southampton, UK}

\begin{document}

\section{Motivations}
 \vskip -0.2cm
  Testing the consistency and the correctness of the Standard Model
 is a central goal of particle physics. The decays of the 
 B-mesons and B-baryons can be used to  
 improve the determination of several poorly
 known matrix elements of the Cabibo-Kobayashi-Maskawa (CKM) quark
 mixing matrix, and in particular $\Vub$.
 The mean values of this
 fundamental parameter of the Standard Model extracted from inclusive 
 decays agree with those extracted from the different exclusive decays
 (such as $\B \rightarrow \pi \ell \nu$ and $\B \rightarrow \tau \nu$) 
 only when the quoted uncertainties are stretched by a factor three 
 \cite{pdg16}. It needs to be resolved whether this arises due to 
 systematic uncertainties inherent in different treatments, 
 or due to beyond Standard Model (BSM) physics.

 We report on the determination of the form
 factors in the $\Bs \rightarrow \K \ell \nu$ decay in $N_f=2$ QCD, 
 and the heavy quark effective theory (HQET) 
 to account for the heavy quark on the lattice \cite{FTB1}. 
 The formalism 
 will be outlined here, especially the 
 non-perturbative renormalization procedure and continuum limit extrapolation. 
 The invariant mass of the leptons 
 is kept fixed as the continuum limit is taken. The differential decay rates
 at a given $q^2$ is related to the renormalized form factors (at 
 the same $q^2$) and $\Vub$.
 A precise experimental determination of this differential decay rate 
 would then allow for a reliable and
 accurate determination of $\Vub$, when all the 
 errors have been systematically accounted for. In our theoretical 
 calculation we identify all the possible systematic 
 sources of error, providing 
 a robust estimate of the exclusive observable. 
 The complementary proceeding \cite{MatK} describes the details involved in the 
 extraction of the bare form factors on the lattice. 

 On the lattice, the most challenging aspect for this reliable computation 
 is the heavy quark itself \cite{RS1}. 
 Due it's mass $m_\mathrm{b} \sim 5 \mathrm{GeV}$ being comparable to that
 of inverse lattice spacing of the finest lattice ensembles in use today,
 the discretization effects due to the heavy quark are particularly 
 severe. A theoretically clean way to avoid this problem is the use of 
 an effective theory for the heavy quark. However, the
 state-of-art computations \cite{Fer1, Flynn, Bouchard, Colquhoun} use 
 either a relativistic heavy quark action or a non-relativistic 
 formulation of QCD on the lattice, which are affected by 
 perturbative renormalization or uncontrolled discretion effects. 
 A fully non-perturbative programme to renormalize the currents does not yet exist. 
 Conventional discretization errors are estimated 
 only by power-counting methods, while the actual extrapolation 
 to the continuum limit with heavy quarks may involve a complicated 
 dependence on the lattice spacing. Our computation seeks to 
 address these issues and demonstrate a clean 
 extrapolation to the continuum limit of the form factors.
 
 \section{Methodology}
 \vskip -0.2cm
 To make contact with phenomenology,  
 we list the following "five-point" program:
 \begin{enumerate}[(i)]
  \setlength{\itemsep}{-3pt}
 \vskip -0.8cm
  \item \emph{reliable} computation of 
  ground state matrix elements $\langle \K | V^\mu(0) | \Bs \rangle$,  
  \item \emph{renormalization} of the matrix elements, either in full QCD, or 
  if an effective theory is being used, then in the effective theory, which is
  then matched to QCD, 
  \item \emph{extrapolation} of the renormalized quantity at finite lattice spacings  
   \emph{to the continuum limit}, 
  \item \emph{extrapolation} of the (light) quark masses to their \emph{physical values},
  \item \emph{mapping} out the $q^2$ dependence of the form factors, since 
  experiments measure values of the differential cross-section.
 \end{enumerate}
 \vskip -0.4cm
  We employ lattice discretized HQET to compute the form factors on the lattice. 
  While the first challenge is the subject of \cite{MatK}, this proceeding is
  concerned with the second and third items. Work on the fourth point is in 
  progress. At the moment, our computation focuses on a fixed value of $q^2$.
  
 \section{HQET on the lattice and non-perturbative renormalization}
 \vskip -0.2cm
  A $\Bs$ state with a momentum $p_{\Bs}$ decays into a Kaon $\K$ 
  with momentum $p_\K$ mediated by the vector current,
  $V^\mu (x) =  \bar{\psi}_\uq (x) \gamma_\mu \psi_\bq (x)$. 
  In QCD, we parameterize the matrix element into two 
  equivalent form-factor decompositions:
  \vspace{-0.1cm}
  \beqa
  \hspace{-0.1cm} \nonumber
  \langle \K (p_\K) | V^\mu(0) | B_s (p_{\Bs}) \rangle &=& 
  \left( p_{\Bs} + p_\K - \frac{m^2_{\Bs} - m^2_\K}{q^2} q\right)^\mu \cdot f_+ (q^2) +
    \frac{m^2_{\Bs} - m^2_\K}{q^2} q^\mu \cdot f_0 (q^2) \\
   &=& \sqrt{2 m_{\Bs}} \left[ v^\mu \cdot h_\parallel (p_\K \cdot v) 
   + p^\mu_\perp \cdot h_\perp (p_\K \cdot v) \right]
  \vspace{-0.1cm}
  \eeqa
   The last line can be seen as a definition of the form factors $h_\parallel$ and $h_\perp$ 
  which we compute. The velocity, momentum and squared momentum transfer ($q^2$) variables 
  are related as: $v^\mu = p^\mu_{\Bs}/m_{\Bs}$, 
  $~p^\mu_\perp = p^\mu_\K - (v \cdot p_\K) v^\mu$,
  $~q^\mu = p^\mu_{\Bs} - p^\mu_\K$, $~p_\K \cdot v = \frac{m^2_{\Bs} + m^2_\K - q^2}{2 m_{\Bs}}$.
    Using non-relativistic state normalization in HQET removes the mass dependence
  present in the relativistic normalization:
  $\langle \Bs (p^\prime) | \Bs (p) \rangle = 2 E({\bf p}) (2\pi)^3 \delta({\bf p} - {\bf p}^\prime)$.
  Consequently, $h_\perp, h_\parallel$ are independent of the heavy quark mass, modulo 
  a logarithmic dependence coming from the matching function between QCD and
  HQET. Note that in the latter, the effective heavy quark fields are mass independent, 
  and hence no additional dependence originates from the vector current. 

  The lattice computation is performed in the rest frame of 
  the $\Bs$ meson: $v^\mu = (1,0,0,0)$, where the QCD matrix 
  elements are simply related to the form factors via:
  \vspace{-0.2cm}
  \beqa 
    (2 m_{\Bs})^{-1/2} \langle  \K (p_\K) | V^0(0) | \Bs (p_{\Bs})  \rangle &=& h_\parallel (E_\K) \\
    (2 m_{\Bs})^{-1/2} \langle  \K (p_\K) | V^k(0) | \Bs (p_{\Bs})  \rangle &=& p^k_\K h_\perp (E_\K) 
  \vspace{-0.2cm}
  \eeqa 
   The only kinematic variable is $E_\K = p_\K \cdot v$, the energy of the 
  final state Kaon. Neglecting terms of $\mathcal{O}(m^2_\ell / m^2_{\Bs}, 
  m^2_\ell / q^2 )$, the differential decay rate can be directly related 
  to the form factor:
  \vspace{-0.2cm}
  \beq
   \frac{\mathrm{d \Gamma} (\Bs \rightarrow \K \ell \nu )}{\mathrm{d}q^2} = 
   \frac{G_\mathrm{F}^2}{24 \pi^3} \Vub^2 |{\bf p}_K|^3 [f_+ (q^2)]^2
  \vspace{-0.2cm}
  \eeq
  With the experimental measurement of the differential decay rate and a 
  reliable estimate of $f_+ (q^2)$, $\Vub$ can be accurately measured. 
  We now discuss the renormalized form factors $h^{\sRGI}_{\parallel, \perp}$, 
  related to the QCD form factors $h_{\parallel,\perp}$ via the matching function(s):
  \beqa
  \vspace{-0.2cm}
   h_{\parallel,\perp} (E_\K) &=& C_{\Vo,\Vk} (M_\bq/ \Lambda_{\MSb})
                     h^{\sRGI}_{\parallel,\perp} (E_\K) \cdot [1 + \mathcal{O}(1/m_\bq)] 
  \vspace{-0.2cm}
  \eeqa

  Our computations use the $N_f=2$ QCD ensembles. A parallel
  project is non-perturbatively matching the currents $V_{0,k}$ between QCD and 
  HQET, and will allow for the $1/m$ corrections to be computed \cite{JH}. For
  now, we remain with the static order, where most of the non-perturbative 
  results are available. The bare currents $V_{0,k}^{\stat}$ are:
  \vspace{-0.1cm}
  \beq
   \Vo^{\mathrm{stat}} = \bar{\psi}_\uq \gamma_0 \psi_\hq + a c_{\Vo} (g_0) \bar{\psi}_\uq 
       \sum_l  \overleftarrow{\nabla}^S \gamma_l \psi_\hq;~~~
   \Vk^{\mathrm{stat}} = \bar{\psi}_\uq \gamma_k \psi_\hq - a c_{\Vk} (g_0) \bar{\psi}_\uq 
       \sum_l  \overleftarrow{\nabla}^S \gamma_l \gamma_k \psi_\hq. 
  \vspace{-0.1cm}
  \eeq
  The leading term is the same as in QCD, except that the heavy quark fields are
  denoted by $\bar{\psi}_\hq$, and the HYP1 and HYP2 heavy quark actions 
  are considered \cite{HYP}, which have an exponentially better 
  signal-to-noise ratio as compared to the classic Eichten-Hill action 
  for the heavy quarks. In addition, there is the $\mathcal{O}(a)$
  improvement term, whose coefficients are currently known to 1-loop order: 
  $c_x = c_x^{(1)} g_0^2 + O(g_0^4)$. They are relatively small \cite{cx}: for HYP1, 
  $c_{\Vo} = 0.0223(6),~~c_{\Vk} = 0.0029(2)$ while for HYP2,
  $c_{\Vo} = 0.0518(2),~~c_{\Vk} = 0.0380(6)$. 
 
  In HQET, the currents get renormalized multiplicatively. At the static order,
  the spin symmetry of HQET results in the same renormalization of the vector current 
  as the axial current: $Z^{\sRGI}_{\Vk} = Z^{\sRGI}_{\Ao}$, but an extra factor 
  $Z^{\stat}_{\mathrm{V/A}} (g_0)$ for $\Vo$, as Wilson fermions break chiral symmetry:
  \vspace{-0.2cm}
  \beq
    \Vo^{\sRGI} = Z^{\sRGI}_{\Ao} (g_0) Z^{\stat}_{\mathrm{V/A}} (g_0) \Vo^{\stat}; 
 ~~~\Vk^{\sRGI} = Z^{\sRGI}_{\Ao} (g_0) \Vk^{\stat}.
  \vspace{-0.2cm}
  \eeq
  The crucial aspect of our non-perturbative renormalization setup proceeds via the  
  calculation of the RGI (renormalization group invariant) quantities $\Phi^{\mathrm{RGI}}$,
  which are both scheme and scale independent. 
  For a renormalized static heavy-light current, the  
  the differential equation for the physical quantity at different scales $\mu$, 
  can be integrated in perturbation theory:
  \vspace{-0.2cm}
  \beq
    (A^\RGI)_\mathrm{0} = \lim_{\mu \to \infty} [2b_0 \bar{g}^2(\mu)]^{-\gamma_0/ 2b_0}
    (A_\mathrm{R}^{\stat})_\mathrm{0} (\mu)
  \vspace{-0.2cm}
  \eeq
  The integration constant in the left hand side is completely independent of scale of 
  the running coupling $\bar{g}^2(\mu)$, the renormalized current and the scheme used to
  compute them. $\gamma_0$ and $b_0$ are the universal 1-loop coefficients of the 
  running mass and the couplings. The goal therefore is to determine the integration constant.
  The non-perturbative analog is expressed as:
  \vspace{-0.2cm}
  \beq
    Z^{\stat}_{\mathrm{A},\RGI} (g_0) = \frac{\Phi^\RGI}{\Phi(\mu)} \times 
     \left. Z^{\stat}_\mathrm{A} (g_0,a\mu) \right|_{\mu=\frac{1}{2 L_\mathrm{max}}}.
  \vspace{-0.2cm}
  \eeq
  The first universal factor relates the renormalization of $A_0^\stat$ at a 
  scale $\mu_0=1/L_\mathrm{max}$ calculated in a scheme to the RGI operator, 
  while the second factor knows about the lattice discretization. 
  The computation of both factors was done in \cite{MDM}, where the Schr\"{o}dinger functional
  scheme was used to compute the RGI quantity, and yielded 
  $\Phi^\RGI/\Phi(\mu)=0.880(7)$. Finally, for our computation, we used a 
  a generous range $[Z^\stat_{V/A}(g_0)]^{-1} = 0.97(3)$, motivated by comparing the
  value for the quenched approximation, noting the absence of $N_f$ dependence
  at the 1-loop order. This uncertainly only affects the $1/m_\hq$
  terms, and will be eliminated in the non-perturbative matching program
  \cite{JH}.

 \section{Matching to QCD}
  \vspace{-0.2cm}
  The renormalized form factors need to be matched to that of QCD
  via the conversion functions $C_x$. These 
  can be obtained by calculating a matrix element upto 
  a given order in both theories and then matching the expressions. 
  As a pedagogical example, consider the 
  matrix element $\mathcal{M}$ of the renormalized axial current in both 1-loop 
  QCD and HQET in the leading order (stat). 
  To make the theories agree, we match the expressions \cite{RS1} obtained for QCD,
  with that of HQET: 
  \vspace{-0.2cm}
  \beqa  
   \mathcal{M}_\mathrm{QCD} (L,m_\hq) &=& [1 + g^2 (-\gamma_0 \ln(m_\hq L) + B_\mathrm{QCD})] 
  \mathcal{M}^{(0)} + \mathcal{O}(g^4) + \mathcal{O}(\frac{1}{m_\hq L}), \\ 
  \mathcal{M}_\mathrm{stat} (\mu L) &=& [1 + g^2 (-\gamma_0 \ln(\mu L) + B_\mathrm{lat})] 
  \mathcal{M}^{(0)} +  \mathcal{O}(g^4).
  \vspace{-0.2cm}
  \eeqa 
  Expressed in terms of the running coupling $g^2$, and renormalized mass, 
  equating the leading order gives the multiplicative matching function between the two
  theories: $C_\mathrm{match} = 1 + g^2 \gamma_0 \ln (\mu/m_\hq) 
  + (B_\mathrm{QCD} - B_\mathrm{lat}) + \mathcal{O}(g^4)$, motivating the  
  the logarithmic dependence on the heavy quark mass.

  A non-perturbative matching of the currents is underway by the ALPHA collaboration
  \cite{JH}. Meanwhile, for our purposes, we use the best available perturbative results.
  A major advantage in our procedure of using the RGI operators is 
  that we can directly use the results from renormalized continuum perturbation 
  theory \cite{Bekavic}, as opposed to other results in the literature 
  which use bare perturbation theory. The limitation here is on the 
  knowledge of the running quark mass, which have uncertainties
  of $\mathcal{O}(\alpha_s^3)$, and translates to the same level of 
  uncertainties on our estimate of the total $Z_x = C_x \times Z_x^{\sRGI}$
  \cite{RS1}. This is already better by one more order in $\alpha_s$ 
  than the other approaches used in the literature which have uncertainties of 
  at least $\mathcal{O}(\alpha_s^2)$. 

 \section{The continuum limit}
  We now outline our results for the continuum limits of the form factors. 
 Computations of the bare form factors are repeated
 for different lattice ensembles with decreasing lattice spacings. 
 For this, the $N_f=2$ ensembles by the Coordinated Lattice Simulations (CLS) 
 effort \cite{CLS} were used. The chosen ensembles (with labels A5, F6 and N6) 
 all have roughly the same pion mass (330, 310, 340 MeV respectively), and 
 the lattice spacing decreases like 0.0749(8), 0.0652(6)
 and 0.0483(4) fm respectively. The spatial dimension of the lattices
 satisfy $m_\pi L > 4$ and an aspect ratio of 2. 
 The ensembles have degenerate light quarks.
 The strange quark mass is fixed by fixing the Kaon mass in units of the Kaon
 decay constant to it's physical value at our light quark masses.

 \begin{figure}
  \begin{center}
   \includegraphics[scale=0.55]{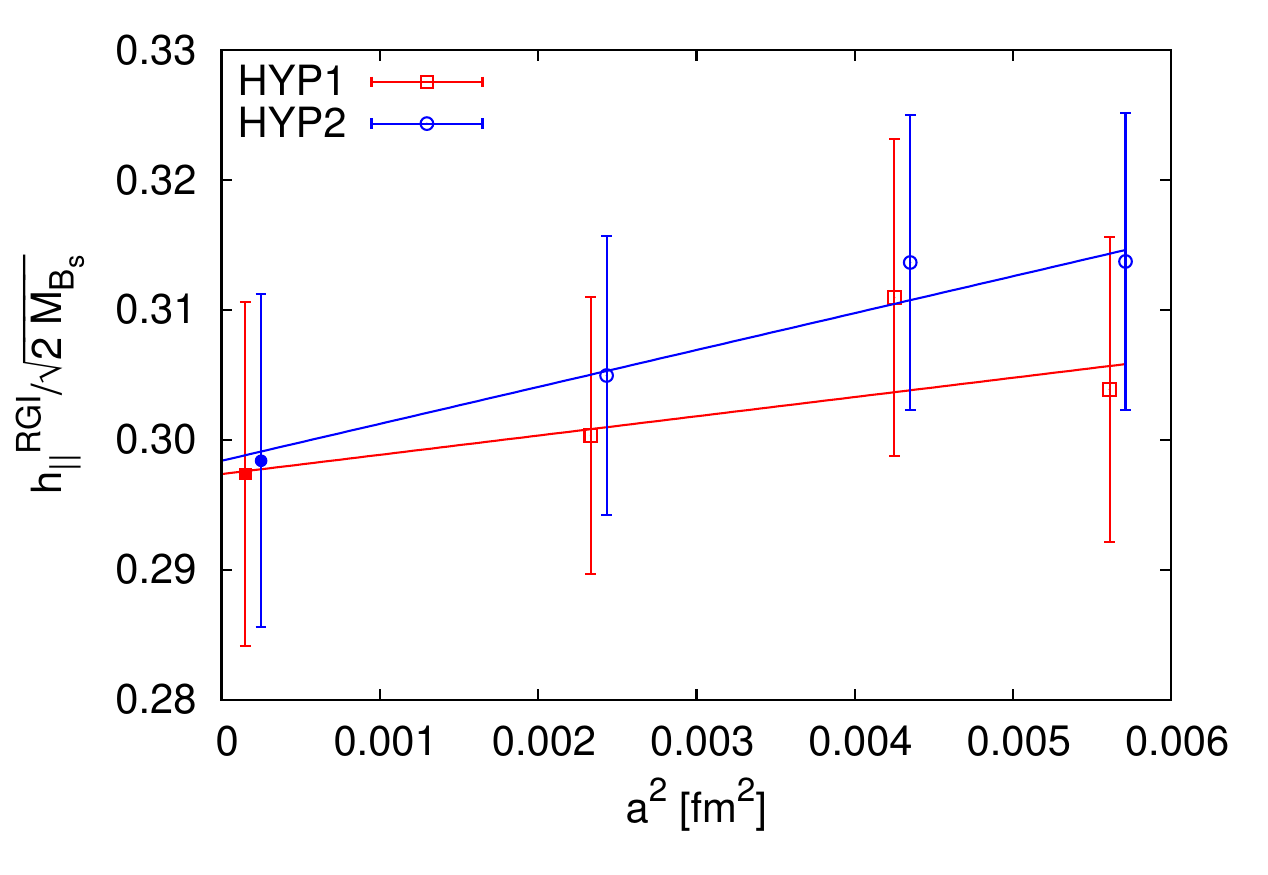}
   \includegraphics[scale=0.55]{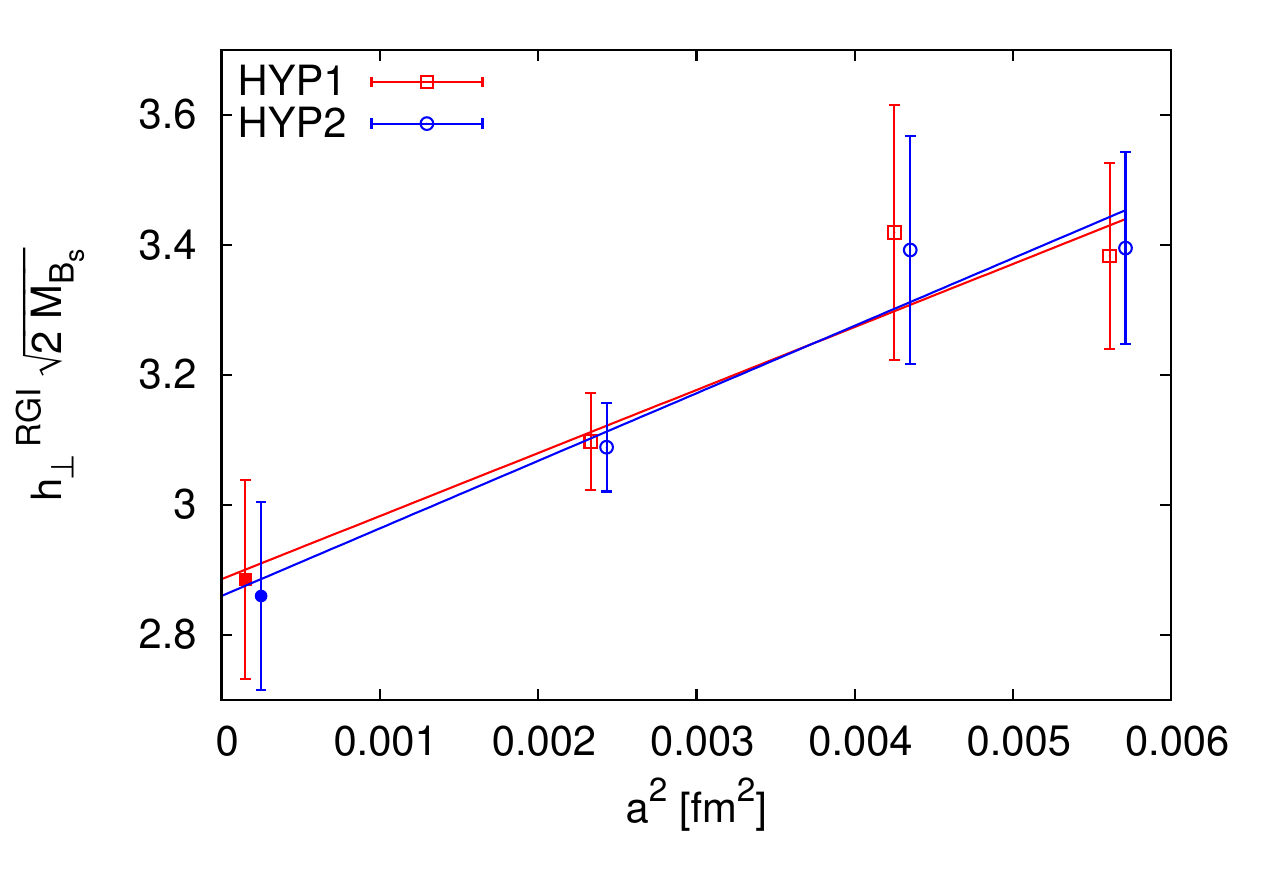}
  \end{center}
   \caption{The linear extrapolation in $a^2$ to the continuum limit 
    of the 1-loop $\mathcal{O}(a)$ improved RGI form factors 
    $h^{\sRGI}_\parallel$ (left) and $h^{\sRGI}_\perp$ (right). 
    The different discretizations HYP1/2 are
    shifted to ensure visibility.} 
   \label{contl}
 \end{figure}
  Since the computations keep the physical momentum ${\bf p}_\K = 0.535$ GeV
 fixed, flavor twisted boundary conditions are necessary to impart the same
 (three-) momentum to the Kaon as the lattice spacing varies: 
 $\psi_\mathrm{s} (x + L \hat{1} ) = e^{i \theta} \psi_\mathrm{s} (x)$. 
 The aforementioned  Kaon momentum is obtained with a lattice 
 momentum of $(1,0,0)$ on the finest lattice (N6). On the F6 and A5, 
 therefore, we choose flavor twisted boundary conditions, 
 such that, ${\bf p}_\K = (1,0,0) (2 \pi + \theta)/L$.
 The F6 gets a larger ($\theta/(2\pi) = 0.350$) twist as compared to A5
 ($\theta/(2\pi) = 0.034$) due to the respective lattice size.
 The $\Bs$ meson is always kept at rest.
 For our value of ${\bf p}_\K$, the momentum transferred is 
 $q^2 = 21.22 ~\mathrm{GeV}^2$ on all lattices, with an error of 
 $0.03-0.05 ~\mathrm{GeV}^2$ due to the lattice spacing. 

 While the extraction of the bare form factors is explained in \cite{MatK},
 we show how the continuum limit can be extracted with the RGI 
 form factors. In the error analysis, the statistical correlations
 and autocorrelations are all taken into account \cite{HQET2, UWerr}.   
 The results of the continuum extrapolation is shown in Fig \ref{contl}.
 A list of the values can be found in \cite{FTB1}. 
 The continuum limit is obtained with an extrapolation linear in $a^2$, with the 
 $c_x = c^{(1)}_x g_0^2$. While the $h_\parallel$ gives a compatible estimate
 by fitting it to a constant (and naturally with smaller error bars), we keep
 the $a^2$ extrapolations because there is no reason why these should 
 disappear. Also, the $\mathcal{O}(a)$ corrections from perturbation theory
 do not matter at the precision of our data. This gives us:
 $h^\sRGI_\parallel = 0.976 (41) \mathrm{GeV}^{1/2}$ and 
 $h^\sRGI_\perp = 0.876 (43) \mathrm{GeV}^{-1/2}$.

  We can now estimate the form factor $f_+$ using
 different systematics for $1/m_\hq$ suppressed terms. Two such estimates are
 $f_{+,1}$ and $f_{+,2}$. In the first one, all the known terms
 and kinematic factors are controlled, expect in $h^\sRGI_{\perp,\parallel}$ 
 (which have suppressed uncontrolled $1/m_\hq$ contributions), while 
 in the second one $f_{+,2}$ all the $1/m_\hq$ terms are systematically dropped: 
 \beqa
  f_{+,1} &=& \sqrt{\frac{m_{\Bs}}{2}} \left(  (1 - \frac{E_\K}{m_{\Bs}}) C_{\Vk}
      h^\sRGI_{\perp} (E_\K) + \frac{1}{m_{\Bs}} C_{\Vo} h^\sRGI_\parallel (E_\K)\right)\\
  f_{+,2} &=& \sqrt{\frac{m_{\Bs}}{2}} C_{\Vk} h^\sRGI_\perp (E_\K).
 \eeqa
 Numerically, the values are: $f_{+,1} = 1.77(7)(7)$ and $f_{+,2} = 1.63(8)(6)$.
 Comparing these estimates, we see that the $\mathcal{O}(1/m_{\hq}$) terms contribute like
 $(1 - \frac{E_\K}{m_{\Bs}})$, and we have a systematic $\sim 15 \%$ 
 ambiguity/uncertainty due to the dropping of these terms. 
 Thus we have a preliminary estimate: 
 $f_+ (21.22 \mathrm{GeV}^2) = f_{+,2} \pm 0.15 f_{+,2} = 1.63(8)(6) \pm 0.24$.
 The second error bars are the $\mathcal{O} (\alpha_s^3)$ 
 perturbative uncertainties in the matching functions.
 We note that this systematic error is expected to drop to $\sim 1-2\%$ 
 when all the $1/m_{\hq}$ will be systematically included.

 The quantity $f_0$ is also often quoted in the literature, which is defined as:
 \beq
 f_0 = \frac{\sqrt{2/m_{\Bs}}} {1 - m^2_\K/m^2_{\Bs}}
 \left(   (1 - \frac{E_\K}{m_{\Bs}}) C_{\Vo} h^\sRGI_\parallel (E_\K) 
    + \frac{{\bf p}^2_\K}{m_{\Bs}} C_{\Vk} h^\sRGI_\perp (E_\K)\right)
 \eeq
 For this number, we estimate $f_0 = 0.663(3)(1)$. The results for $f_{+,0}$
 agree well with those existing in the literature: For Flynn et al. \cite{Flynn},
 the form factors extracted at our values of $q^2$ are $f_+ \simeq 1.65(10)$ and
 $f_0 \simeq 0.62(5)$, while Bouchard et al. \cite{Bouchard} report  
 $f_+ \simeq 1.80(20)$ and $f_0 \simeq 0.66(5)$. Given that our calculation has
 a very different source of systematic uncertainty, this
 agreement is very necessary for phenomenological applications. 

 \section{Conclusion and Outlook}
  We report the first study of the continuum limit of non-perturbatively
 renormalized form factors. Since the stress is on the continuum limit
 we have kept the momentum transfer fixed at $q^2 = 21.22 \mathrm{GeV}^2$. 
 In the static limit, we have precise non-perturbative determinations of 
 the RGI form factors. The matching to QCD is perturbative, 
 and only has uncertainties of order $\mathcal{O} (\alpha_s^3)$. 
 The discretization effects are not very large, allowing a 
 smooth linear $a^2$ extrapolation to the continuum. On the other hand 
 $1/m_\hq$ effects need to incorporated.

  The agreement of the various methods in the extraction of the form factors
 increase our confidence in their extraction using lattice techniques. Once the
 $1/m_\hq$ are included they will be of direct phenomenological interest. 
 This is envisaged as a direct follow-up of the project. Preliminary investigations
 are already in progress. For connection to phenomenology, the
 extrapolation to the physical light quarks will also be considered, as well as
 the computation at another value of the momentum. The basic results
 for the continuum extrapolation presented here inspire the confidence that
 precise lattice results, can be used together with experimental measurements
 for a reliable extraction of $V_\mathrm{ub}$ in the near future.

\end{document}